\documentclass[%
 reprint,
 amsmath,amssymb,
 aps,
]{revtex4-1}

\usepackage{graphicx}% Include figure files
\usepackage{dcolumn}% Align table columns on decimal point
\usepackage{bm}% bold math
\begin{document}

\preprint{APS/123-QED}

\title{Topology of DNA: a honeycomb stable structure under salt effect}% Force line breaks with \\

\author{Elsa de la Calleja}
%\author{}%
\affiliation{%
 Instituto de Investigaciones en Materiales, Universidad Nacional Aut\'onoma de M\'exico, Apdo. Postal 70-360, Ciudad Universitaria, 04510, M\'exico.\\
}%


\author{R. F Bazoni, M. S. Rocha}
\affiliation{
 Departamento de F\'{i}sica, Universidade Federal de Vicosa (UFV), Av. P. H. Rolfs s/n. CEP 36570-900. Vicosa - MG. Brazil.\\
}%
%\affiliation{
 %Third institution, the second for Charlie Author
%}%
\author{Marcia Barbosa}
\affiliation{%
 Instituto de F\'{i}sica, Universidade Federal do Rio Grande do Sul, Caixa Postal 15051, 91501-970, Porto Alegre, RS, Brazil\\
}%



\begin{abstract}
Atomic Force Microscopy analysis is employed to study the geometrical and topological properties of $3000$kbp DNA molecules fixed in mica substrates with $MgCl_{2}$. We found that the aggregates on the substrate surface for certain salt concentrations form a honeycomb stable structure with the addition of salt. The honeycomb, and the transition to other structures, were characterized by the \emph{Betti numbers}, which is a topological invariant property and by \emph{Hausdorff-Besicovitch fractal dimension} 
\begin{description}
\item[PACS numbers]
May be entered using the \verb+\pacs{#1}+ command.
\end{description}
\end{abstract}

\pacs{Valid PACS appear here}% PACS, the Physics and Astronomy
\maketitle

%\tableofcontents

\section{Introduction}
The macromolecule of DNA is composed by phosphate groups linked to the base-pairs which form a quiral double helix~\cite{Franklin:1953dg}. In the presence of water, the phosphate groups are dissociated, cations are released in the solution and the isolated DNA forms an elongated relatively rigid structure~\cite{Pasos:2014dg}. With the addition of monovalent salt the excess of ions screen the electrostatic interactions and the DNA becomes more flexible~\cite{Bte:1996dg}. In a more dense solution of DNA, the macromolecule-macromolecule interaction becomes relevant and the scenario is more complex~\cite{Pasos:2014dg,Rivetti:1996dg,Livro:2011dg}. The DNA-salt interaction leads to the formation of structures, including  aggregates with a honeycomb topology~\cite{Kee:2012dg,Singh:2016dg,Chen:1995dg,Young-Sang:2016dg,Burak:2003dg,Wissenburg:1995dg}.

The honeycomb topology is ubiquitous in nature and has been found in a number of physical and biological structures. In addition it has been also explored in mathematical simple models such as in the ideal lattices~\cite{Owerre:2016dg} due to the potential use in  designing and creating network-on-chip~\cite{Xu:2009dg}, or as inspiration for new engineering structures~\cite{Zhang:2015dg}.

The macromolecule of DNA has been studied widely by different techniques ~\cite{Bte:1996dg,Bu:1996dg,Bezanilla:1993dg,Bust:1996dg,Rivetti:1996dg} and length scales~\cite{Bu:1996dg,Livro:2011dg,Israelachvili:2011dg}. It has been demonstrated that the intrinsic curvature of DNA can be changed by the addition of low concentrations of chloroquine and ethidium bromide~\cite{Kee:2012dg} or the entropic elasticity is affected by intercalating in the 
DNA structure other molecules~\cite{Marcio:2007dg}. The relation between changes in the DNA structure and flexibility and the addition of salt is confirmed in a number of experiments. The fluorescence microscopy shows that in the presence of salt a honeycomb-like topology~\cite{Singh:2016dg} is observed. The electron microscopy and the gel-electrophoresis experiments show that the kinetoplast DNA presents a honeycomb structure when the solution contains $100$mM of [NaCl]~\cite{Chen:1995dg}, however the cyclic tetrameric and pentameric isomers have not been identified~\cite{Chen:1995dg}. Atomic Force Microscopy (AFM) and fluorescence microscopy measurements found that a variety of DNA networks can be formed on the $SiO_2$ substrate by controlling the buffer solution and substrate surface~\cite{Young-Sang:2016dg}. Even though the topology and salt concentration seems to be related, no numerical characterization of this relationship exists.

In this paper we show by Atomic Force Microscopy experiments that the addition of salt changes the topology of the dsDNA aggregates. This topological transition is then measured by the \emph{Betti number} approach which  provides a theoretical framework for explaining the mechanism behind the dsDNA aggregation process. 

%%%%%%%%%%%%%%%%%%%%%%%%%%%%%%%%%%%%%%%%
\section{Materials and Methods}
%%%%%%%%%%%%%%%%%%%%%%%%%%%%%%%%%%%%%%%%
%%%%%%%%%%%%%%%%%%%%%%%%%%%%%%%%%%%%%%%%
\subsection{Sample Preparation}
%%%%%%%%%%%%%%%%%%%%%%%%%%%%%%%%%%%%%%%%
We follow a standard procedure to prepare the samples, where DNA is diluted in a solution with different salt concentrations. The solution of deposition contains $7.5\mu M$ of $3kbp$ of DNA molecule, plus TRIS ($10mM$) + HCl buffer with a $pH=7.4$, and $5mM$ of $MgCl_{2}$ necessary to fix the DNA molecules on mica substrates~\cite{Rivetti:1996dg,Young-Sang:2016dg}. Different [NaCl] concentrations were added to the solution for each sample. The salt concentration used were: $3mM$, $7mM$, $20mM$, $100mM$ and $150mM$.

The waiting time for the deposition was $15$ minutes. After the solution was deposited on the surface of a cleaved mica, the  solution is kept still  for $10$ minutes and then the sample was washed with DI-water to remove the excess of the salt excess and the crystals. Finally, the surface was dried with a weak nitrogen jet, after the excess of water was removed with a special tissue. We use Atomic Force Microscopy to characterize the structural effects of the  [NaCl] on DNA. We employed a $2nm$ diameter tip with $\sim 270 Hz$ as the resonance frequency in tapping mode, to avoid damage to our sample. Several AFM images of the DNA solution were obtained for each salt concentration. Based on these image, the global behavior of adsorbed polymers on mica.

%%%%%%%%%%%%%%%%%%%%%%%%%%%%%%%%%%%%%%%%
\subsection{The Multifractal Spectrum}
%%%%%%%%%%%%%%%%%%%%%%%%%%%%%%%%%%%%%%%%
The standard box counting method was employed to measure the generalized \emph{Hausdorff-Besicovitch fractal dimension} of two dimensional images of the final stages of DNA's aggregates for each salt concentration. The geometrical description by means of the  fractal dimension was employed successfully in many physical systems ~\cite{Barnsley:1993dg,Benoit:1977dg,Vicsek:1989dg,Meakin:1988dg}  including the DNA~ \cite{Cattani:2013dg,Voss:1992adg}.

The procedure to measure the fractal dimension goes as follows: the image is binarized by a high contrast treatment, leaving the particles  black and the space between them white. A grid of four random positions covers the entire image with a decreasing size of $\varepsilon$ as the length of the box. The scaling law to relate the number of particles and the size of the boxes follow the relation $N(\varepsilon)\sim\varepsilon^{-D_{Q}}$, where $\varepsilon$ acquired successively smaller values of length until the minimum value of $\varepsilon_{0}$ and $N(\varepsilon)$ are the number of cubes required to coverall the set. The fractal dimension by the box counting method is given by~\cite{Halsey:1986dg,Hentschel:1983dg,Procaccia:1983dg,Ott:1993dg,Feigenbaum:1986dg,HalseyErr:1986dg}
%%%%%%%%%%%%%%%%%%%%%
\begin{equation}
D_{Q}=\lim_{\varepsilon \rightarrow 0}\frac{\ln N(\varepsilon)}{\ln (\varepsilon_{0}/\varepsilon)}
\end{equation}
%%%%%%%%%%%%
Then the statistical properties were described by the local fractal dimension by means of the  generalized box counting method
 defined as
%%%%%%%%%%%%%%%%%%%%
\begin{equation}
D_{Q}=\frac{1}{1-q}\lim_{\varepsilon\rightarrow 0}\frac{ln I(Q,\varepsilon)}{ln (\varepsilon_{0}/\varepsilon)}
\end{equation}
%%%%%%%%%%%%%%%%%
where
%%%%%%%%%%%%%%
\begin{equation}
I(q,\varepsilon)=\sum_{i=1}^{N(\varepsilon)}[P_{(i,Q)}]^{Q}
\end{equation}
%%%%%%%%%%%%%%%%%%%
We used the scaling exponent defined by Halsey et al.~\cite{Halsey:1986dg,HalseyErr:1986dg} as $P_{i,q}^{q}\sim\varepsilon_{i}^{\alpha q}$ where $\alpha$ can take a width range of values measuring different regions of the set. The spectrum generated by an infinite set of dimension $D_{q}=D_{0},D_{1},D_{2},...$ measure the scaling structure as a function of the local pattern density. If \emph{q=0} the generalized fractal dimension represent the classic fractal dimension, it means that $D_{f}=D_{q=0}$. As the image is divided into pieces of size $\varepsilon$, it suggested that the number of times that $\alpha$ in $P_{i,q}$ takes a value between $\alpha'$ and $d\alpha'$  defined as $d\alpha'\rho(\alpha')\varepsilon^{-f(\alpha')}$ where $f(\alpha')$ is a continuous function.
As \emph{q} represents different scaling indices, we can define
%%%%%%%%%%%%%%%%%%%%%%%%%%%%%%%%%%
\begin{equation}
I(q,\varepsilon)=\sum_{i=1}^{N(\varepsilon)}[P_{(i,q)}]^{q}=\int d\alpha'\rho(\alpha')\varepsilon^{-f(\alpha')+q\alpha'}
\end{equation}
%%%%%%%%%%%%%%%%%%%%%%%%%%%%%%%%%%
$\alpha_{i}$ is the  Lipschitz-H\"{o}lder exponent, that characterizes the singularity strength in the \emph{ith} box. The factor $\alpha_{i}$ allows to
quantify the distribution of complexity in an spatial location. The multifractal is a set of overlapping self-similar configurations. In that way, we used the scaling relationship taking into account $f(\alpha)$ as a function to cover a length scales of observations. Defining the number of boxes as a function of the Lipschitz-H\"{o}lder exponent $(\alpha)$, can be related to the box size $\varepsilon$~\cite{Ashvin:1989dg} as 
%%%%%%%%%%%%%%%%%%%%%%%%%%%%%%%%%%
\begin{equation}
N(\alpha)\sim\varepsilon^{-f(\alpha)}
\end{equation}
%%%%%%%%%%%%%%%%%%%%%%%%%%%%%%%%%%
where $f(\alpha)$ is a continuous function~\cite{Procaccia:1983dg,Elsama:2016dg}.

The multifractal spectrum show a line of consecutive points for $Q \geq 0$ that start on the left side of the spectrum climbing up to the maximum value. Then the values for $Q\leq0$, represented in the spectrum for a dotted line, on the right side begins to descend. The maximum value corresponds to $Q=0$, which is equal to the box counting dimension. To obtain the multi-fractal spectrum we use the plugin \emph{FracLac}. Basically $D_{q}$ is the variation of mass as a function of $\varepsilon$ in the image, and give us the behavior as a power series of $\varepsilon$ sizes distorting them by an exponent \emph{q}. We select the case of
%%%%%%%%%%%%%%%%%%%%%%%%
\begin{equation}
D_{f}=D_{q=0}
\end{equation}
as the parameter of order in the images. In the plugin we select four grid positions that cover the total image, and the mode scaled series was selected to see the singularity spectrum results. The final configuration of the simulations, present a particle distribution of particles in black. The parameters of the program were calibrate for this kind of images.

%We used the scaling exponent defined by Halsey et al.~\cite{Halsey:1986dg,HalseyErr:1986dg}  as $P_{i,q}^{q}\sim\varepsilon_{i}^{\alpha q}$. $\alpha_q$ assumes a wide range of values measuring different regions of the set. The spectrum generates an infinite set of dimensions $D_{q}=D_{0},D_{1},D_{2},... $ which measures the scaling structure as a function of the local pattern density. If \emph{Q=0} the generalized fractal dimension represents the classic fractal dimension, $D_{f}=D_{Q=0}$. As the image is divided into pieces of size $\varepsilon$, it is suggested that the number of times that $\alpha$ in $P_{i,Q}$ takes a value between $\alpha'$ and $d\alpha'$  defined as $d\alpha'\rho(\alpha')\varepsilon^{-f(\alpha')}$ where $f(\alpha')$ is a continuous function~\cite{Procaccia:1983dg,Elsama:2016dg}.

%The multifractal exponent is obtained from  set of overlapping self-similar configurations. In that way, we used the scaling relations to obtain $f(\alpha)$ as a function which covers the length scales of observations. Defining the number of boxes as a function of the Lipschitz-H\"{o}lder exponent $N(\alpha)$, is related to the box size $\varepsilon$ as $N(\alpha)\sim\varepsilon^{-f(\alpha)}\cite{Ashvin:1989dg}$. 

%To obtain the multifractal spectrum we use the plugin~\emph{FracLac}\cite{FracLac:2013dg}. Basically $D_{Q}$ is the variation of mass as a function of $\varepsilon$ in the image, and give us the behavior as a power series of $\varepsilon$ sizes distorting them by an exponent \emph{Q}. We select the case of $D_{f}=D_{Q=0}$ as the parameter of order in the images. 

\subsection{Topological invariants}

Homology can be used to distinguish between objects by constructing algebraic invariants that exhibit their connectivity properties. The topological properties such as the number of connected components and the number of $\mu$-dimensional holes, can be used to characterize spatial structures~\cite{Mecke:2002dg}.

The \emph{Betti numbers}, $\beta_{\mu}$, describe in a complete way the topological structure of the system~\cite{Robins:1989dg}. For a two-dimensional geometry or image, two Betti numbers can be computed, $\beta_0$ and $\beta_1$. The first measures the number of connected regions and the second the number of holes that represent the light spaces on the network. The procedure to compute the two numbers goes as follows. The pictures were transformed in a gray level images. Then we process these pictures using the following topological formalism.

If we define a polytope as a subset of $\Re^{d}$ and if a subset $X$ of $\Re^{d}$ is homeomorphic to this polytope, then $X$ is triangulated by a simplicity complex $\zeta$~\cite{Mecke:2002dg,Robins:1989dg}.
The group structure given by $\zeta$ defines the addition of $\mu$-simplistic. The free group resulting, the chain group $\zeta_{\mu}(X)$, has $\mu$-chains, the sum of a finite number of oriented $\mu$-simplistic~\cite{Robins:1989dg}.
Using the boundary operator, we define the image of $\partial_{\mu}$ as a subgroup of the boundary group $B_{\mu-1}$. The set of all $\mu$-chains with empty boundary is the group of $\mu$-cycles, $Z_{\mu}$ defined as the null space of $\partial_{\mu}$.

Given that $H_{\mu}(K)$ is the quotient group in the homology group, then  $H_{\mu}=Z_{\mu}/B_{\mu}$, where $Z_{\mu}(K)$ is the cycle group with finite rank. This means that two $\mu$-cycles $x_{\mu}$ and $y_{\mu}$ belong to the same homology class if the $\mu$-chain formed by their difference is the boundary of some ($\mu$+1)-chain, $y_{\mu}$-$x_{\mu}$=$\partial$$v_{\mu+1}$.

Within this formalism the \emph{Betti numbers} of \emph{X} in the dimension $\mu$ are the topological invariants of $|X|$.~\cite{Munkres:1984dg}.
And the number of distinct equivalence classes of $H_{\mu}$ is the $\mu$th \emph{Betti number} $\beta_{\mu}$.
Then $\mu$th Betti number counts the number of $\mu$-dimensional holes in \emph{X}. When $\mu$=0, the Betti number counts the number of
path-connected components of \emph{X}. For subsets of $\Re^{3}$, we can interpret $\beta_{0}$ as the number of independent lines, and $\beta_{1}$ as the number of enclosed spaces. For a two-dimensional geometry or image, two \emph{Betti number} can be compute $\beta_{0}$ and $\beta_{1}$.

 In order to obtain the \emph{Betti numbers} the images were processed using the ImageJ software~\citep{Schneider:2012dg}. High resolution images were transformed, through standard image processing algorithms, into binary images.Such binary images are amenable for the calculation of the Betti numbers using the software CHOMP\cite{Chomp:2017dg}. Each image was transformed into an 8-bit gray resolution and then the brightness was adjusted to obtain the  dark lines of the structure. The image was then binarized taking into consideration a threshold level of 132 a.u. (see Fig.~\ref{fig:dna}). The binary image is taken to represent a 2-dimensional topological space for which two Betti numbers,
$\beta_{0}$ and $\beta_{1}$, can be calculated.

%%%%%%%%%%%%%%%%%%%%%%%%%%%%%%%%%%%%%%%%%%%%%
\section{Results and Discussions}
%%%%%%%%%%%%%%%%%%%%%%%%%%%%%%%%%%%%%%%%%%%%

%%%%%%%%%%%%%%%%%%%%%%%%%%%%%%%%%%%%%%%%%%%%%
\subsection{AFM of the DNA-salt solution}
%%%%%%%%%%%%%%%%%%%%%%%%%%%%%%%%%%%%%%%%%%%%%%%%%%%%%%%

\begin{figure}[ht]
\centering
\includegraphics[width=4cm]{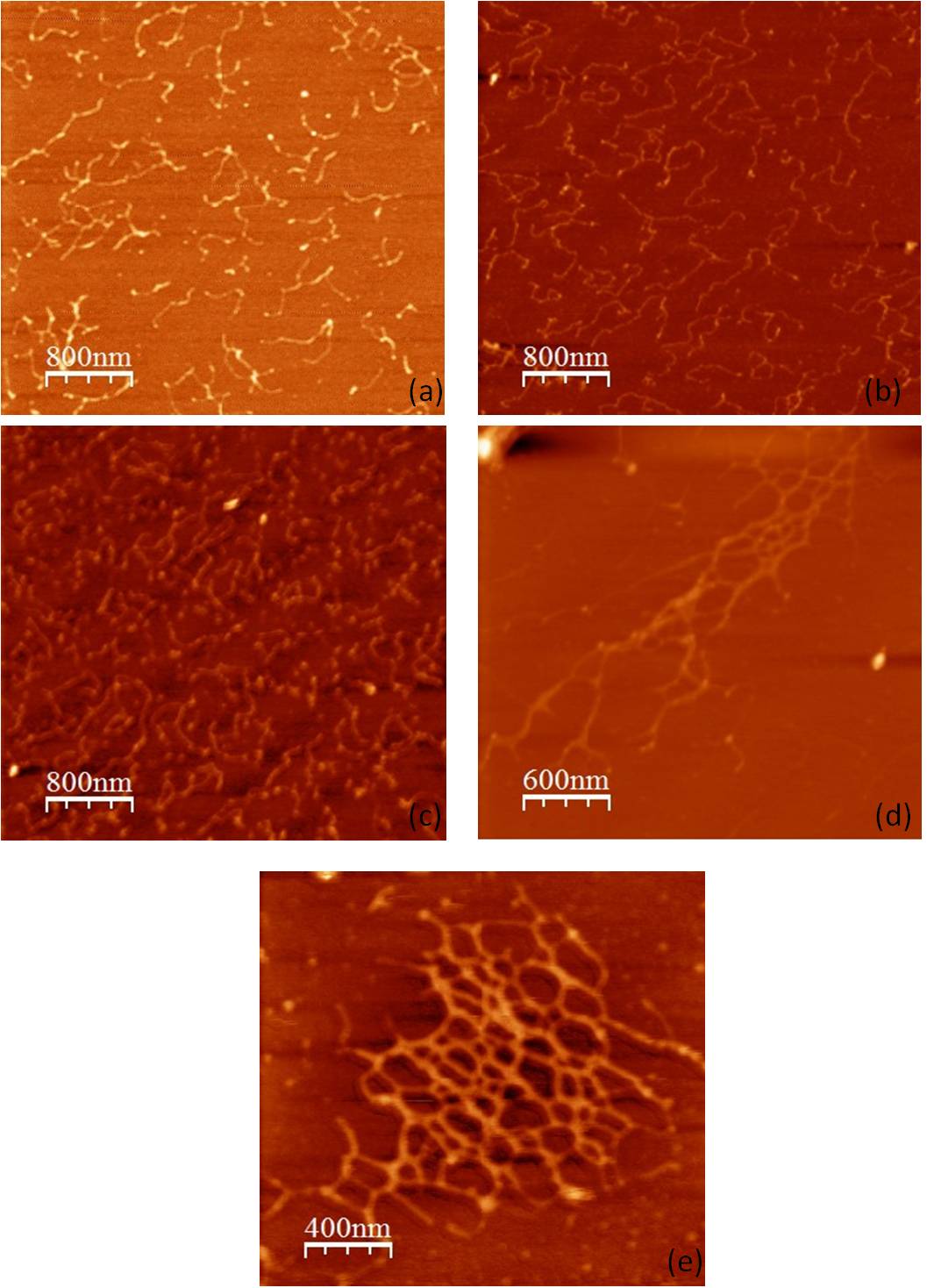}
\caption{The structural evolution of the dsDNA  (a) for $3mM$, (b)  $7mM$ (c)  $20mM$, (d) $100mM$ and (e) $150mM$ of [NaCl] salt concentrations.}
\label{fig:FIG-1}
\end{figure}

Figure~\ref{fig:FIG-1} shows the images obtained by AFM of the dsDNA on a substrate with different [NaCl] concentrations prepared as described in the previous section. At the lowest salt concentration, $3mM$, the macromolecules are isolated forming individual elongated polymers. As the salt concentration increases,  $7mM$, the dsDNA decreases the intermolecular repulsion and the cross-linked polymers are created. For even higher salt concentrations, $20mM$, the cross-linked aggregates are formed. As the salt concentration is increased even further, $100mM$,the  DNA bundles are formed. For the  highest salt concentration, $150mM$, the DNA bundles percolate the system forming a two dimensional honeycomb topology.

In order to develop a methodology to measure the patterns and the transitions between them three assumptions are made. First, we consider only the connected layer on each structure. In the surface, isolated dsDNA and small structures are also observed, but since they are not statistically significantly, they are ignored. Second, we neglect the depth of the structures on the images, which implies that the network is assumed to be a mono-layer. Due to the electrostatic repulsion of the phosphate groups, layering of dsDNA is not expected. Third, we assume that the high brightness represents a high local concentration of [NaCl] which indicates a change of direction on the line of the dsDNA. All these three assumptions are supported by the available experimental data. 
%%%%%%%%%%%%%%%%%%%%%%%%%%%%%%%%%%%%%%%%%%%%%%%%%%%%%%%%%%%%%%%%%%
%%%%%%%%%%%%%%%%%%%%%%%%%%%%%%%%%%%%%%%%%%%%%%%%%%%%%%%%%%%%%%%%%%%%%%%%%%%%%%%%%

%%%%%%%%%%%%%%%%%%%%%%%%%%%%%%%%%%%%%%%%%%%%%
\subsection{Fractal dimension of the DNA-salt solution}
%%%%%%%%%%%%%%%%%%%%%%%%%%%%%%%%%%%%%%%%%%%%%%%%%%%%%%%

\begin{figure}[ht!]
\centering
\includegraphics[width=0.96\linewidth]{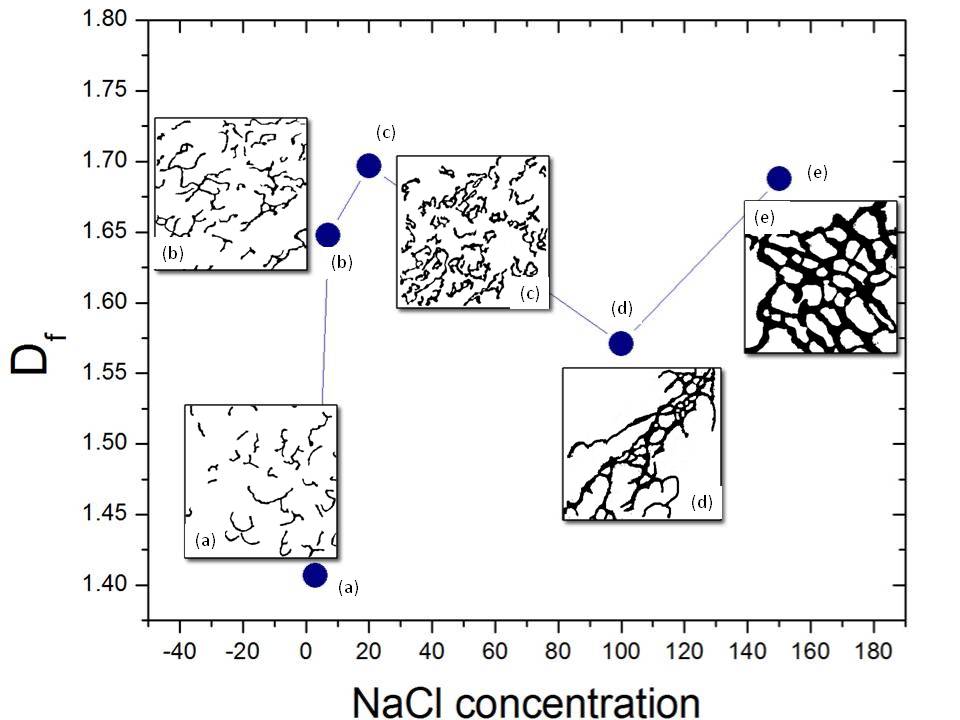}
\caption{Generalized fractal dimension as a function of the salt concentration. The pictures of
the system at each salt concentration are illustrated.}
\label{fig:FDF}
\end{figure}

Figure~\ref{fig:FDF} shows the fractal dimension of the final stages of the DNA, $D_{f}$, as a function of salt concentrations computed by the method described in the previous section.  A picture  of the binarized structured of each stage illustrates the structure formed by DNA, which changes as a function of the [NaCl] concentration. Figure~\ref{fig:FDF}\emph{(a)} corresponds to a picture of the DNA under $3mM$ of [NaCl] salt. The low value of the fractal dimension indicates that the structure is geometrically close to a one-dimensional line. As the salt concentration increases, the fractal dimension also increases approaching the two dimensional value. The electrostatic repulsion is screened and the polymers form cross-links polymers and cross-linked aggregates as illustrates in the photographs \emph{(b)} and \emph{(c)}. For even higher concentrations the system forms bundles due to the electrostatic attraction mediated by the salt~\cite{Singh:2016dg,Young-Sang:2016dg,Chen:1995dg}. The formation of these dense linear structures reflects in the decrease of the fractal dimension. It is consequence of geometrical characteristics of the structure~\cite{Elsama2:2017dg,Falconer:2014dg}. For the highest salt concentration analyzed in this study the bundles percolate the system forming a quasi-two dimensional structure and the fractal dimension increases. This final structure show a honeycomb-type structure. We found some isolated structures on the surface of the mica and it was measured by AFM at $200nm$ of resolution. 

%%%%%%%%%%%%%%%%%%%%%%%%%%%%%%%%%%%%%%%%%%%%%%%%%%%%%%
%%%%%%%%%%%%%%%%%%%%%%%%%%%%%%%%%%%%%%%%%%%%%%%%%%%%%%

The aggregation process induced by ion effect are visibly evident on the presented experimental results shown by AFM images. The aggregation process experimented by DNA with [NaCl] produces different final states with different degrees of fractal dimension. We identify that fractal dimension increases as a function of the complexity of linked-DNA; however, the fractal dimension of an image of a honeycomb is around $D_{f}=1.77$, and the fractal dimension measured on the final structure formed on a salt concentration of $150mM$ was $D_{f}=1.7$ what suggests that this final structure is honeycomb.
Our results indicate that there is an aggregation process of the DNA which is affected by electrostatic interactions~\cite{Stanley:1988dg,Elsama:2010dg}. These structural variations provide information about the macroscopic effects~\cite{Elsama:2016dg,Elsama:2017dg} and they become important tool for understanding biological processes such as transcription and replication.

%It has been discussed the implications of the relation of the dimension of an object and the dimension by the occupied space by the object\cite{Bunde:1991dg,Falconer:2014dg,Barabasi:1995dg,Elsama2:2017dg}. The basic relation between fractal dimension and topological dimension is $D_{f}>D_{T}$, where $D_{T}$ is the topological dimension\cite{Barabasi:1995dg,Elsama2:2017dg,Bunde:1991dg,Falconer:2014dg}. To complete the characterization of the structural changes in the linked-DNA we used topological invariant.

%%%%%%%%%%%%%%%%%%%%%%%%%%%%%%%%%%%%%%%%%%%%%%%%%%%%%%
\subsection{Topological invariants to measure assembly of the DNA-structure}
%%%%%%%%%%%%%%%%%%%%%%%%%%%%%%%%%%%%%%%%%%%%%%%%%%%%%%

\begin{figure}[ht]
\centering
\includegraphics[width=0.96\linewidth]{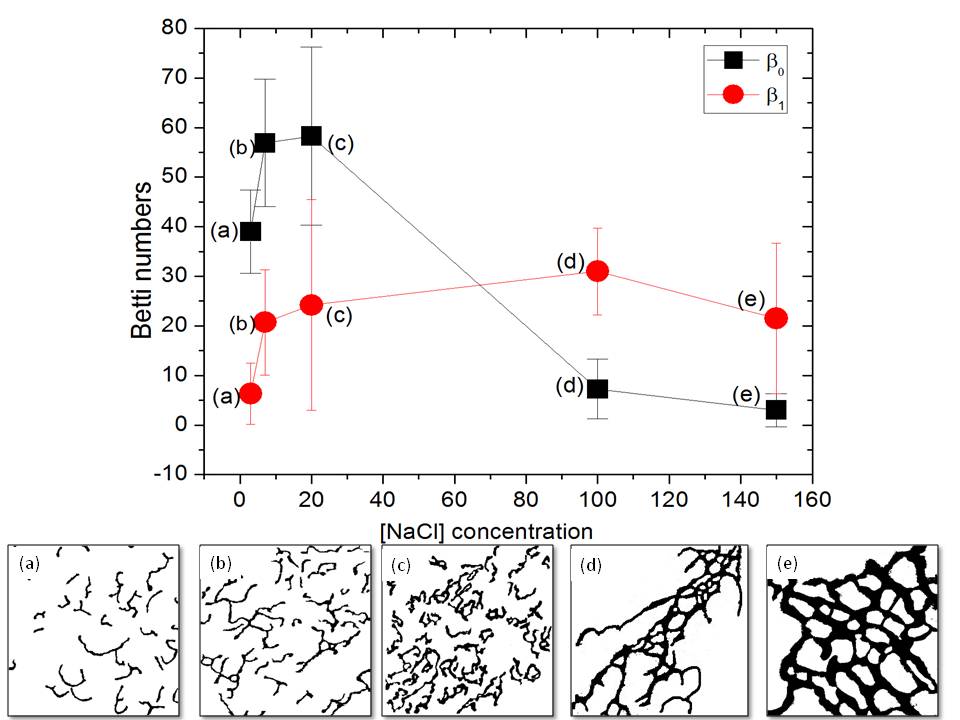}
\caption{ $\beta_{0}$ and $\beta_{1}$ versus salt concentration. At the bottom the photographs of the system a the salt concentrations $3mM$, $7mM$, $20mM$,  $100mM$ and $150mM$.}
\label{fig:FBN}
\end{figure}

The fractal dimension, even though consistent with the photographs and topological transitions observed in the DNA solution, can not provide a concise picture of the system. The fractal dimension of the cross-linked cluster is the same as the fractal dimension of the honeycomb therefore, the calculation of the $D_f$ is unable to distinguish between these two structures. In order to circumvent this difficulty another topological measurement is employed.
We calculate the average of connectivity by \emph{Betti numbers} for the different images obtained by AFM (showed in Figure~\ref{fig:FIG-1}) when dsDNA is submitted to different salt concentrations. The behavior of $\beta_{0}$ gives the information about the topological connectivity, while the $\beta_{1}$ gives the number of holes or empty spaces surrounded by black lines. Figure~\ref{fig:FBN} shows the values of  the \emph{Betti numbers} for all the final stages of the DNA under the chosen salt concentration. The figure also illustrates the binarized pictures corresponding to the each final structure for the different salt concentration used in the experiments. 

Figure~\ref{fig:FBN} illustrates how the topological connectivity, $\beta_{0}$, changes with the addition of salt. For the cases \emph{(a)}, \emph{(b)} and \emph{(c)} the behavior of connectivity follows the behavior of fractal dimension as illustrated in the Figure~\ref{fig:FDF}, increasing with the addition of salt. The relation between $D_{f}$ and $\beta_{0}$ can be explained as a function of the density of black pixels\cite{Elsama2:2017dg}. 
These stages show how the dsDNA are in process of link to each other. 
In the cases \emph{(d)} and \emph{(e)} the system forms one large cluster and the connectivity increases, it means, one structure is almost completely linked.
The behavior $\beta_{0}$ differentiate the cases \emph{(c)} and \emph{(e)} that show similar values for the fractal dimension $D_f$ as illustrated in the Figure~\ref{fig:FDF}. Even though in both cases the surface is covered, for the case \emph{(e)} one structure is formed while in the case \emph{(c)} many structures are present\cite{Elsama2:2017dg}.

Figure~\ref{fig:FBN} also shows the behavior of $\beta_{1}$ which correspond to the number of holes or the empty spaces inside linked DNA. The number of holes increases as the salt concentration increases, because the dsDNA are grouped into structures such as pentagons, hexagons or closed regions, and it increase to finally come to form a honeycomb-like structure.
The topological invariants gives us information about the complexity of the structure\cite{Elsama:2017dg} and we found that there is possible to predict the final structure of DNA under salt effect.
%%%%%%%%%%%%%%%%%%%%%%%%%%%%%%%%%%%%%%%%%%%%%%%%%
%%%%%%%%%%%%%%%%%%%%%%%%%%%%%%%%%%%%%%%%%%%%%%%%%

%%%%%%%%%%%%%%%%%%%%%%%%%%%%%%%%%%%%%%%%%%%%%%%%%
\begin{figure}[ht]
\centering
\includegraphics[width=0.6\linewidth]{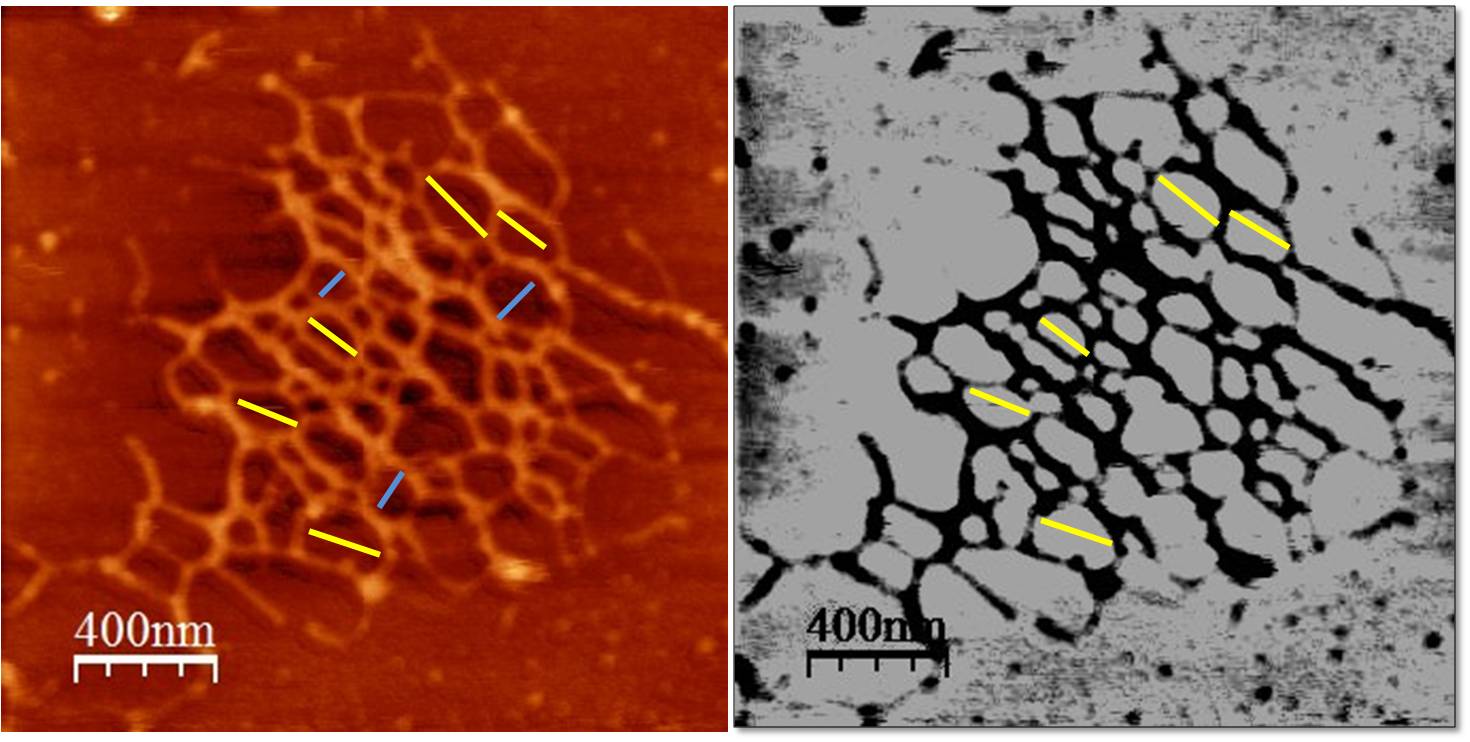}
\caption{The image on the left show an image obtained by AFM of the final structure of DNA under $150mM$ of salt concentration on a scale of $400nm$. On the right is presented the same image after the threshold treatment. The yellow marks correspond to the length of the holes and the blue ones the width.}
\label{fig:dna}
\end{figure}
%%%%%%%%%%%%%%%%%%%%%%%%%%%%%%%%%%%%%%%%%%%%%%%%%
Figure~\ref{fig:dna} show a image of the final structure of DNA where is under $150mM$ of [NaCl]. The yellow marks represents the length of the holes, and the blue ones the width. We found that whether in the $10$ minutes that we left the solution of deposition or in the $15$ minutes that we left for achieve the balance in the solution, the mean length of the holes is $189.78 nm$, and the mean width is $108nm$ of the hexagons on the honeycomb structure. However is possible to see a high dispersion of sizes.

Since the beginning of the process we identify the tendency to form pentagons and hexagons as a function of salt concentration. However, despite the evolution of the topological structure is not simple, we did not identity addition of pentagons and hexagons in a simple way. The complex evolution of DNA under salt effect, represents a rearrangement of the electrostatic interactions on the phosphate groups, and the final effect of salt on DNA was to form a kind of honeycomb topology.

There has been reported that DNA is a semi-flexible polymer, this characteristic affect directly the possible options that can be formed when dsDNA is affected by [NaCl]. The flexible properties of DNA helps explain why the dsDNA does not strain the individual rings\cite{Rauch:1991dg}, and could be the reason that we observed that there is a persistent length that seems like this is keeping on the honeycomb structure. It seems that, DNA aggregates in a honeycomb structure naturally, following the principle of better accommodation \cite{Hales:2005dg}

%%%%%%%%%%%%%%%%%%%%%%%%%%%%%%%%%%%%%%%%%%%%%%%%%
\section{Conclusions}
%%%%%%%%%%%%%%%%%%%%%%%%%%%%%%%%%%%%%%%%%%%%%%%%%
In this paper we described the aggregation process of DNA submitted to different [NaCl] concentrations. At low salt concentrations the  $3000kbp$ DNA due to the rigid hydrogen bonds and non screened  electrostatic repulsion form isolated DNA structures. As the salt concentration increases the screening of both the electrostatic and hydrogen bonds interactions takes place. Two mechanisms take place. First, the repulsion between the different segments one single DNA is screened as the salt concentration  increases and the DNA becomes more flexible. Then, the DNA-DNA repulsion decreases with more salt and a network of the macromolecules is formed. The combination of these two these two phenomena leads to a sequence of structural changes illustrated by our experimental results. 
The \emph{Hausdorff-Besicovitch} fractal dimension show the increase of $D_f$ with salt with a final value that corresponds to the values of a honeycomb structure.  $\beta_{1}$, which measures the number of holes,
also shows the transition from isolated extended  DNA to more entangled structure promoted by the increase of the flexibility of the DNA.  The $\beta_0$, which measures the connectivity, shows the effect of the increase of salt screening the DNA-DNA repulsion and promoting the formation of the network. Our results suggest that the DN honeycomb structure is a result  of this two steps process.

%%%REFERENCES%%%
\bibliography{rsc} %You need to replace "rsc" on this line with the name of your .bib file

\providecommand*{\mcitethebibliography}{\thebibliography}
\csname @ifundefined\endcsname{endmcitethebibliography}
{\let\endmcitethebibliography\endthebibliography}{}
\begin{mcitethebibliography}{45}
\providecommand*{\natexlab}[1]{#1}
\providecommand*{\mciteSetBstSublistMode}[1]{}
\providecommand*{\mciteSetBstMaxWidthForm}[2]{}
\providecommand*{\mciteBstWouldAddEndPuncttrue}
  {\def\EndOfBibitem{\unskip.}}
\providecommand*{\mciteBstWouldAddEndPunctfalse}
  {\let\EndOfBibitem\relax}
\providecommand*{\mciteSetBstMidEndSepPunct}[3]{}
\providecommand*{\mciteSetBstSublistLabelBeginEnd}[3]{}
\providecommand*{\EndOfBibitem}{}
\mciteSetBstSublistMode{f}
\mciteSetBstMaxWidthForm{subitem}
{(\emph{\alph{mcitesubitemcount}})}
\mciteSetBstSublistLabelBeginEnd{\mcitemaxwidthsubitemform\space}
{\relax}{\relax}

\bibitem[Rosalind and Gossing(1953)]{Franklin:1953dg}
E.~F. Rosalind and R.~G. Gossing, \emph{Nature}, 1953, \textbf{171},
  740--741\relax
\mciteBstWouldAddEndPuncttrue
\mciteSetBstMidEndSepPunct{\mcitedefaultmidpunct}
{\mcitedefaultendpunct}{\mcitedefaultseppunct}\relax
\EndOfBibitem
\bibitem[Passos and Barbosa(2014)]{Pasos:2014dg}
C.~B. Passos and M.~C. Barbosa, \emph{Physica A}, 2014, \textbf{413}, 481\relax
\mciteBstWouldAddEndPuncttrue
\mciteSetBstMidEndSepPunct{\mcitedefaultmidpunct}
{\mcitedefaultendpunct}{\mcitedefaultseppunct}\relax
\EndOfBibitem
\bibitem[Smith \emph{et~al.}(1996)Smith, Cui, and Bustamante]{Bte:1996dg}
S.~B. Smith, Y.~Cui and C.~Bustamante, \emph{Science}, 1996, \textbf{271},
  795\relax
\mciteBstWouldAddEndPuncttrue
\mciteSetBstMidEndSepPunct{\mcitedefaultmidpunct}
{\mcitedefaultendpunct}{\mcitedefaultseppunct}\relax
\EndOfBibitem
\bibitem[Rivetti \emph{et~al.}(1996)Rivetti, Guthold, and
  Bustamante]{Rivetti:1996dg}
C.~Rivetti, M.~Guthold and C.~Bustamante, \emph{J. of Molecular Biology.},
  1996, \textbf{264}, 919–932\relax
\mciteBstWouldAddEndPuncttrue
\mciteSetBstMidEndSepPunct{\mcitedefaultmidpunct}
{\mcitedefaultendpunct}{\mcitedefaultseppunct}\relax
\EndOfBibitem
\bibitem[Braga and Ricci(2004)]{Livro:2011dg}
P.~C. Braga and D.~Ricci, \emph{Atomic Force Microscopy in Biomedical
  Research.}, 2004, vol.~1, p. 515\relax
\mciteBstWouldAddEndPuncttrue
\mciteSetBstMidEndSepPunct{\mcitedefaultmidpunct}
{\mcitedefaultendpunct}{\mcitedefaultseppunct}\relax
\EndOfBibitem
\bibitem[Tan \emph{et~al.}(2012)Tan, Li, Gray, Yang, Tao-Ng, Zhang, Rui-Tan,
  Hiew, Lee, and Li]{Kee:2012dg}
H.~K. Tan, D.~Li, R.~K. Gray, Z.~Yang, M.~T. Tao-Ng, H.~Zhang, J.~M. Rui-Tan,
  S.~H. Hiew, J.~Y. Lee and T.~Li, \emph{Org. Biomol. Chem.}, 2012,
  \textbf{10}, 2227--2230\relax
\mciteBstWouldAddEndPuncttrue
\mciteSetBstMidEndSepPunct{\mcitedefaultmidpunct}
{\mcitedefaultendpunct}{\mcitedefaultseppunct}\relax
\EndOfBibitem
\bibitem[Sushma \emph{et~al.}(2016)Sushma, Singh, Kumbhar, and
  Khan]{Singh:2016dg}
B.~Sushma, A.~Singh, S.~Kumbhar and A.~Khan, \emph{Chem. Eur. J.}, 2016,
  \textbf{22}, 1--13\relax
\mciteBstWouldAddEndPuncttrue
\mciteSetBstMidEndSepPunct{\mcitedefaultmidpunct}
{\mcitedefaultendpunct}{\mcitedefaultseppunct}\relax
\EndOfBibitem
\bibitem[Chen \emph{et~al.}(1995)Chen, Rauch, White, and
  Cozzarelli]{Chen:1995dg}
J.~Chen, C.~A. Rauch, P.~T. White, J.~H.~Englund and N.~R. Cozzarelli,
  \emph{Cell Press}, 1995, \textbf{80}, 61--69\relax
\mciteBstWouldAddEndPuncttrue
\mciteSetBstMidEndSepPunct{\mcitedefaultmidpunct}
{\mcitedefaultendpunct}{\mcitedefaultseppunct}\relax
\EndOfBibitem
\bibitem[Young-Sang~Jo(2003)]{Young-Sang:2016dg}
Y.~R. Young-Sang~Jo, Younghun~Lee, \emph{Materials Science and Engineering C.},
  2003, \textbf{23}, 851–855\relax
\mciteBstWouldAddEndPuncttrue
\mciteSetBstMidEndSepPunct{\mcitedefaultmidpunct}
{\mcitedefaultendpunct}{\mcitedefaultseppunct}\relax
\EndOfBibitem
\bibitem[Burak \emph{et~al.}(2003)Burak, Ariel, and Andelman]{Burak:2003dg}
Y.~Burak, G.~Ariel and D.~Andelman, \emph{Biophysical Journal}, 2003,
  \textbf{85}, 2100--2110\relax
\mciteBstWouldAddEndPuncttrue
\mciteSetBstMidEndSepPunct{\mcitedefaultmidpunct}
{\mcitedefaultendpunct}{\mcitedefaultseppunct}\relax
\EndOfBibitem
\bibitem[Wissenburg \emph{et~al.}(1995)Wissenburg, Odijk, Cirkel, and
  Mandel]{Wissenburg:1995dg}
P.~Wissenburg, T.~Odijk, P.~Cirkel and M.~Mandel, \emph{Macromolecules}, 1995,
  \textbf{28}, 2315--2328\relax
\mciteBstWouldAddEndPuncttrue
\mciteSetBstMidEndSepPunct{\mcitedefaultmidpunct}
{\mcitedefaultendpunct}{\mcitedefaultseppunct}\relax
\EndOfBibitem
\bibitem[Owerre(2016)]{Owerre:2016dg}
S.~A. Owerre, \emph{Journal of Physics: Condensed Matter}, 2016, \textbf{28},
  38\relax
\mciteBstWouldAddEndPuncttrue
\mciteSetBstMidEndSepPunct{\mcitedefaultmidpunct}
{\mcitedefaultendpunct}{\mcitedefaultseppunct}\relax
\EndOfBibitem
\bibitem[Xu \emph{et~al.}(2009)Xu, Wei-Yin, H., and P.]{Xu:2009dg}
T.~C. Xu, A.~Wei-Yin, T.~H. and L.~P., \emph{Network and Parallel Computing
  Workshops, IFIP International Conference on}, 2009, \textbf{0}, 73--79\relax
\mciteBstWouldAddEndPuncttrue
\mciteSetBstMidEndSepPunct{\mcitedefaultmidpunct}
{\mcitedefaultendpunct}{\mcitedefaultseppunct}\relax
\EndOfBibitem
\bibitem[Zhang \emph{et~al.}(2015)Zhang, Yang, Li, Huang, Feng, Shen, Han,
  Zhang, Jin, Xu, and Lu]{Zhang:2015dg}
Q.~Zhang, X.~Yang, P.~Li, G.~Huang, S.~Feng, C.~Shen, B.~Han, X.~Zhang, F.~Jin,
  F.~Xu and T.~J. Lu, \emph{Progress in Materials Science}, 2015, \textbf{74},
  332--400\relax
\mciteBstWouldAddEndPuncttrue
\mciteSetBstMidEndSepPunct{\mcitedefaultmidpunct}
{\mcitedefaultendpunct}{\mcitedefaultseppunct}\relax
\EndOfBibitem
\bibitem[Bustamante and C.(1996)]{Bu:1996dg}
C.~Bustamante and R.~C., \emph{Ann. Rev. Biophys. Biomol. Struc.}, 1996,
  \textbf{25}, 395--429\relax
\mciteBstWouldAddEndPuncttrue
\mciteSetBstMidEndSepPunct{\mcitedefaultmidpunct}
{\mcitedefaultendpunct}{\mcitedefaultseppunct}\relax
\EndOfBibitem
\bibitem[Bezanilla \emph{et~al.}(1993)Bezanilla, Bustamante, and
  G.]{Bezanilla:1993dg}
M.~Bezanilla, C.~Bustamante and H.~H. G., \emph{Scanning Microscopy}, 1993,
  \textbf{7(4)}, 1145--1148\relax
\mciteBstWouldAddEndPuncttrue
\mciteSetBstMidEndSepPunct{\mcitedefaultmidpunct}
{\mcitedefaultendpunct}{\mcitedefaultseppunct}\relax
\EndOfBibitem
\bibitem[Garcia and Reich(1996)]{Bust:1996dg}
C.~Garcia, R.~A.~Bustamante and N.~O. Reich, \emph{Proc. Natl. Acad.}, 1996,
  \textbf{Sci. USA 93}, 7618--7622\relax
\mciteBstWouldAddEndPuncttrue
\mciteSetBstMidEndSepPunct{\mcitedefaultmidpunct}
{\mcitedefaultendpunct}{\mcitedefaultseppunct}\relax
\EndOfBibitem
\bibitem[Israelachvili(2011)]{Israelachvili:2011dg}
J.~N. Israelachvili, \emph{Intermolecular and Surfaces Forces.}, 2011, vol.
  Third Edition, USA\relax
\mciteBstWouldAddEndPuncttrue
\mciteSetBstMidEndSepPunct{\mcitedefaultmidpunct}
{\mcitedefaultendpunct}{\mcitedefaultseppunct}\relax
\EndOfBibitem
\bibitem[Rocha \emph{et~al.}(2007)Rocha, Ferreira, and Mesquita]{Marcio:2007dg}
M.~S. Rocha, M.~C. Ferreira and O.~N. Mesquita, \emph{The Journal of Chemical
  Physics}, 2007, \textbf{127}, 105108\relax
\mciteBstWouldAddEndPuncttrue
\mciteSetBstMidEndSepPunct{\mcitedefaultmidpunct}
{\mcitedefaultendpunct}{\mcitedefaultseppunct}\relax
\EndOfBibitem
\bibitem[Barnsley(1993)]{Barnsley:1993dg}
M.~F. Barnsley, \emph{Fractals Everywhere.}, 1993, vol.~1, p. 531\relax
\mciteBstWouldAddEndPuncttrue
\mciteSetBstMidEndSepPunct{\mcitedefaultmidpunct}
{\mcitedefaultendpunct}{\mcitedefaultseppunct}\relax
\EndOfBibitem
\bibitem[Mandelbrot(1977)]{Benoit:1977dg}
B.~Mandelbrot, \emph{The Fractal Geometry of Nature.}, 1977, vol.~1, p.
  468\relax
\mciteBstWouldAddEndPuncttrue
\mciteSetBstMidEndSepPunct{\mcitedefaultmidpunct}
{\mcitedefaultendpunct}{\mcitedefaultseppunct}\relax
\EndOfBibitem
\bibitem[Vicsek(1989)]{Vicsek:1989dg}
T.~Vicsek, \emph{Fractal Growth Phenomena.}, 1989, vol.~1, p. 488\relax
\mciteBstWouldAddEndPuncttrue
\mciteSetBstMidEndSepPunct{\mcitedefaultmidpunct}
{\mcitedefaultendpunct}{\mcitedefaultseppunct}\relax
\EndOfBibitem
\bibitem[Meakin(1988a)]{Meakin:1988dg}
P.~Meakin, \emph{The growth of fractal aggregates and their fractal measures,
  in Phase Transitions and Critical Phenomena.}, 1988a, vol.~12, p.
  336–489\relax
\mciteBstWouldAddEndPuncttrue
\mciteSetBstMidEndSepPunct{\mcitedefaultmidpunct}
{\mcitedefaultendpunct}{\mcitedefaultseppunct}\relax
\EndOfBibitem
\bibitem[Cattani and Pierro(2013)]{Cattani:2013dg}
C.~Cattani and G.~Pierro, \emph{Bull Math Biol}, 2013, \textbf{75}, 1544\relax
\mciteBstWouldAddEndPuncttrue
\mciteSetBstMidEndSepPunct{\mcitedefaultmidpunct}
{\mcitedefaultendpunct}{\mcitedefaultseppunct}\relax
\EndOfBibitem
\bibitem[Voss(1992)]{Voss:1992adg}
R.~F. Voss, \emph{Phys. Rev. Lett.}, 1992, \textbf{68(25)}, 3805–3808\relax
\mciteBstWouldAddEndPuncttrue
\mciteSetBstMidEndSepPunct{\mcitedefaultmidpunct}
{\mcitedefaultendpunct}{\mcitedefaultseppunct}\relax
\EndOfBibitem
\bibitem[Halsey \emph{et~al.}(1986)Halsey, Jensen, Kadanoff, Procaccia, and
  Shraiman]{Halsey:1986dg}
T.~C. Halsey, M.~H. Jensen, L.~P. Kadanoff, I.~Procaccia and B.~I. Shraiman,
  \emph{Phys. Rev. A}, 1986, \textbf{33}, 1141\relax
\mciteBstWouldAddEndPuncttrue
\mciteSetBstMidEndSepPunct{\mcitedefaultmidpunct}
{\mcitedefaultendpunct}{\mcitedefaultseppunct}\relax
\EndOfBibitem
\bibitem[Hentschel and Procaccia(1983)]{Hentschel:1983dg}
G.~E. Hentschel and I.~Procaccia, \emph{Physica D: Nonlinear Phenomena}, 1983,
  \textbf{8}, 435--444\relax
\mciteBstWouldAddEndPuncttrue
\mciteSetBstMidEndSepPunct{\mcitedefaultmidpunct}
{\mcitedefaultendpunct}{\mcitedefaultseppunct}\relax
\EndOfBibitem
\bibitem[Hentschel and Procaccia(1983)]{Procaccia:1983dg}
H.~G.~E. Hentschel and I.~Procaccia, \emph{Physica D}, 1983, \textbf{8},
  435\relax
\mciteBstWouldAddEndPuncttrue
\mciteSetBstMidEndSepPunct{\mcitedefaultmidpunct}
{\mcitedefaultendpunct}{\mcitedefaultseppunct}\relax
\EndOfBibitem
\bibitem[Ott(1993)]{Ott:1993dg}
E.~Ott, \emph{Chaos in Dynamical Systems.}, 1993, vol.~1, p. 385\relax
\mciteBstWouldAddEndPuncttrue
\mciteSetBstMidEndSepPunct{\mcitedefaultmidpunct}
{\mcitedefaultendpunct}{\mcitedefaultseppunct}\relax
\EndOfBibitem
\bibitem[Feigenbaum \emph{et~al.}(1986)Feigenbaum, Jensen, and
  Procaccia]{Feigenbaum:1986dg}
M.~J. Feigenbaum, M.~H. Jensen and I.~Procaccia, \emph{Phys. Rev. Lett.}, 1986,
  \textbf{57}, 1503\relax
\mciteBstWouldAddEndPuncttrue
\mciteSetBstMidEndSepPunct{\mcitedefaultmidpunct}
{\mcitedefaultendpunct}{\mcitedefaultseppunct}\relax
\EndOfBibitem
\bibitem[Halsey \emph{et~al.}(1986)Halsey, Jensen, Kadanoff, Procaccia, and
  Shraiman]{HalseyErr:1986dg}
T.~C. Halsey, M.~H. Jensen, L.~P. Kadanoff, I.~Procaccia and B.~I. Shraiman,
  \emph{Phys. Rev. A}, 1986, \textbf{34}, 1601\relax
\mciteBstWouldAddEndPuncttrue
\mciteSetBstMidEndSepPunct{\mcitedefaultmidpunct}
{\mcitedefaultendpunct}{\mcitedefaultseppunct}\relax
\EndOfBibitem
\bibitem[Chhabra \emph{et~al.}(1989)Chhabra, Meneveau, Jensen, and
  Sreenivasan]{Ashvin:1989dg}
A.~B. Chhabra, C.~Meneveau, R.~V. Jensen and K.~R. Sreenivasan, \emph{Phys.
  Rev. A}, 1989, \textbf{40}, 5284\relax
\mciteBstWouldAddEndPuncttrue
\mciteSetBstMidEndSepPunct{\mcitedefaultmidpunct}
{\mcitedefaultendpunct}{\mcitedefaultseppunct}\relax
\EndOfBibitem
\bibitem[de~la Calleja \emph{et~al.}(2016)de~la Calleja, Cervantes, and de~la
  Calleja]{Elsama:2016dg}
E.~M. de~la Calleja, F.~Cervantes and J.~de~la Calleja, \emph{Annals of
  Physics}, 2016, \textbf{371}, 313--322\relax
\mciteBstWouldAddEndPuncttrue
\mciteSetBstMidEndSepPunct{\mcitedefaultmidpunct}
{\mcitedefaultendpunct}{\mcitedefaultseppunct}\relax
\EndOfBibitem
\bibitem[Mecke and Stoyan(2008)]{Mecke:2002dg}
K.~R. Mecke and D.~Stoyan, \emph{Morphology of Condensed Matter: Physics and
  Geometry of Spatially Complex Systems.}, 2008, vol.~1, p. 442\relax
\mciteBstWouldAddEndPuncttrue
\mciteSetBstMidEndSepPunct{\mcitedefaultmidpunct}
{\mcitedefaultendpunct}{\mcitedefaultseppunct}\relax
\EndOfBibitem
\bibitem[Robins(2002)]{Robins:1989dg}
V.~Robins, \emph{Computational topology for point data: Betti numbers of
  $\alpha$-shapes.}, 2002, vol. 80(1), pp. 261--274\relax
\mciteBstWouldAddEndPuncttrue
\mciteSetBstMidEndSepPunct{\mcitedefaultmidpunct}
{\mcitedefaultendpunct}{\mcitedefaultseppunct}\relax
\EndOfBibitem
\bibitem[Munkres(19845)]{Munkres:1984dg}
J.~R. Munkres, \emph{Elements of Algebraic Topology}, Addison-Wesley Publishing
  Co., 1st edn., 19845\relax
\mciteBstWouldAddEndPuncttrue
\mciteSetBstMidEndSepPunct{\mcitedefaultmidpunct}
{\mcitedefaultendpunct}{\mcitedefaultseppunct}\relax
\EndOfBibitem
\bibitem[Schneider \emph{et~al.}(2012)Schneider, Rasband, and
  Eliceiri]{Schneider:2012dg}
C.~Schneider, W.~Rasband and K.~Eliceiri, \emph{Nature Methods}, 2012,
  \textbf{9}, 671--675\relax
\mciteBstWouldAddEndPuncttrue
\mciteSetBstMidEndSepPunct{\mcitedefaultmidpunct}
{\mcitedefaultendpunct}{\mcitedefaultseppunct}\relax
\EndOfBibitem
\bibitem[homology project(2017)]{Chomp:2017dg}
C.~homology project, 2017\relax
\mciteBstWouldAddEndPuncttrue
\mciteSetBstMidEndSepPunct{\mcitedefaultmidpunct}
{\mcitedefaultendpunct}{\mcitedefaultseppunct}\relax
\EndOfBibitem
\bibitem[de~la Calleja and Zenit(2017)]{Elsama2:2017dg}
E.~M. de~la Calleja and R.~Zenit, \emph{unpublished}, 2017,  1--15\relax
\mciteBstWouldAddEndPuncttrue
\mciteSetBstMidEndSepPunct{\mcitedefaultmidpunct}
{\mcitedefaultendpunct}{\mcitedefaultseppunct}\relax
\EndOfBibitem
\bibitem[Falconer(2014)]{Falconer:2014dg}
K.~Falconer, \emph{Fractal geometry. Mathematical foundations and
  applications}, 2014, vol.~1, p. 288\relax
\mciteBstWouldAddEndPuncttrue
\mciteSetBstMidEndSepPunct{\mcitedefaultmidpunct}
{\mcitedefaultendpunct}{\mcitedefaultseppunct}\relax
\EndOfBibitem
\bibitem[Stanley and Meakin(1988)]{Stanley:1988dg}
H.~E. Stanley and P.~Meakin, \emph{Nature}, 1988, \textbf{335}, 405--409\relax
\mciteBstWouldAddEndPuncttrue
\mciteSetBstMidEndSepPunct{\mcitedefaultmidpunct}
{\mcitedefaultendpunct}{\mcitedefaultseppunct}\relax
\EndOfBibitem
\bibitem[de~la Calleja \emph{et~al.}(2013)de~la Calleja, Carrillo, Mendoza, and
  Donado]{Elsama:2010dg}
E.~M. de~la Calleja, J.~L. Carrillo, M.~E. Mendoza and F.~Donado, \emph{The
  European Physical Journal B}, 2013, \textbf{86}, 126\relax
\mciteBstWouldAddEndPuncttrue
\mciteSetBstMidEndSepPunct{\mcitedefaultmidpunct}
{\mcitedefaultendpunct}{\mcitedefaultseppunct}\relax
\EndOfBibitem
\bibitem[de~la Calleja and Zenit(2017)]{Elsama:2017dg}
E.~M. de~la Calleja and R.~Zenit, \emph{Knowledge-Based Systems}, 2017,
  \textbf{126}, 48--55\relax
\mciteBstWouldAddEndPuncttrue
\mciteSetBstMidEndSepPunct{\mcitedefaultmidpunct}
{\mcitedefaultendpunct}{\mcitedefaultseppunct}\relax
\EndOfBibitem
\bibitem[Chen \emph{et~al.}(1995)Chen, Rauch, White, Englund, and
  Cozzarelli]{Rauch:1991dg}
J.~Chen, C.~A. Rauch, J.~H. White, P.~T. Englund and N.~R. Cozzarelli,
  \emph{Cell}, 1995, \textbf{80(1)}, 61--69\relax
\mciteBstWouldAddEndPuncttrue
\mciteSetBstMidEndSepPunct{\mcitedefaultmidpunct}
{\mcitedefaultendpunct}{\mcitedefaultseppunct}\relax
\EndOfBibitem
\bibitem[Hales(2005)]{Hales:2005dg}
T.~C. Hales, \emph{Annals of Mathematics}, 2005, \textbf{162}, 1065--1185\relax
\mciteBstWouldAddEndPuncttrue
\mciteSetBstMidEndSepPunct{\mcitedefaultmidpunct}
{\mcitedefaultendpunct}{\mcitedefaultseppunct}\relax
\EndOfBibitem
\end{mcitethebibliography}
\bibliographystyle{rsc}

\end{document}